\documentclass[10pt,conference]{IEEEtran}

\usepackage{amssymb}
\usepackage{amsfonts}
\usepackage{amsmath}
\usepackage{latexsym}
\usepackage{epsfig}
\usepackage{psfrag}
\usepackage{bm}
\usepackage{xspace}
\usepackage{graphicx}
\usepackage{cite}

\renewcommand{\baselinestretch}{0.94}

\newcommand{\defeq}{\triangleq}

\newcommand{\ignore}[1]{}

\newcommand{\R}{\mathbb{R}}

\newcommand{\PG}[2]{\mathrm{PG}(#1,#2)}
\newcommand{\EG}[2]{\mathrm{EG}(#1,#2)}

\newcommand{\PGq}{\operatorname{PG}(2,q)}
\newcommand{\EGq}{\operatorname{EG}(2,q)}

\newcommand{\code}[1]{\mathcal{#1}}
\newcommand{\codePG}[2]{\code{C}_{\PG{#1}{#2}}}
\newcommand{\codeEG}[2]{\code{C}_{\EG{#1}{#2}}}
\newcommand{\codePGq}{\code{C}_{\PGq}}
\newcommand{\codeEGq}{\code{C}_{\EGq}}

\newcommand{\set}[1]{\mathcal{#1}}

\newcommand{\matr}[1]{\mathbf{#1}}
\newcommand{\vect}[1]{\mathbf{#1}}

\newcommand{\dmins}{d_{\mathrm{min}}}

\newcommand{\wH}{w_{\mathrm{H}}}
\newcommand{\wHmin}{w_{\mathrm{H}}^{\mathrm{min}}}

\newcommand{\wpsAWGNC}{w_{\mathrm{p}}^{\mathrm{AWGNC}}}

\newcommand{\wcol}{w_{\mathrm{col}}}
\newcommand{\wrow}{w_{\mathrm{row}}}

\newcommand{\tr}{\mathsf{T}}

\newcommand{\vx}{\vect{x}}
\newcommand{\hvx}{\vect{\hat x}}
\newcommand{\vy}{\vect{y}}

\newcommand{\vlambda}{\boldsymbol{\lambda}}
\newcommand{\vomega}{\boldsymbol{\omega}}

\newcommand{\convhull}{\operatorname{conv}}

\newcommand{\supp}{\operatorname{supp}}

\newcommand{\fp}[1]{\set{#1}}
\newcommand{\fph}[2]{\set{#1}(\matr{#2})}

\newcommand{\fch}[2]{\set{#1}(\matr{#2})}

\newcommand{\setDML}[1]{\set{D}^{\mathrm{ML}}_{#1}}
\newcommand{\setDMLzero}{\set{D}^{\mathrm{ML}}_{\vect{0}}}
\newcommand{\setDLPzero}{\set{D}^{\mathrm{LP}}_{\vect{0}}}

\newcommand{\Mps}{\set{M}_{\mathrm{p}}}

\newcommand{\chicw}[1]{\chi_{#1}^{\mathrm{CW}}}
\newcommand{\chimcw}[1]{\chi_{#1}^{\mathrm{MCW}}}
\newcommand{\chiAWGNCmpcw}[1]{\chi_{#1}^{\mathrm{MPCW}, \mathrm{AWGNC}}}

\newtheorem{Definition}{Definition}

\newtheorem{Theorem}[Definition]{Theorem}

  {\noindent \emph{Proof:}}{\hfill$\square$}

\newcommand{\etheorem}{\hfill$\square$}
\newcommand{\edefinition}{\hfill$\square$}

\begin{document}

\title{On the Minimal Pseudo-Codewords of \\ Codes 
       from Finite Geometries}

\author{
\authorblockN{Pascal O.~Vontobel\authorrefmark{1},
              Roxana Smarandache\authorrefmark{2},
              Negar Kiyavash\authorrefmark{3},
              Jason Teutsch\authorrefmark{4},
              Dejan Vukobratovic\authorrefmark{5}}
\authorblockA{\authorrefmark{1}
                 Dept.~of ECE,
                 University of Wisconsin,
                 Madison, WI 53706, USA,
                 \texttt{vontobel@ece.wisc.edu}}
\authorblockA{\authorrefmark{2}
                 Dept.~of Math.~and Stat.,
                 San Diego State University,
                 San Diego, CA 92182, USA, 
                 \texttt{rsmarand@math.sdsu.edu}}
\authorblockA{\authorrefmark{3}
                 CSL and Dept.~of ECE,
                 University of Illinois,
                 Urbana, IL 61801, USA,
                 \texttt{kiyavash@uiuc.edu}}
\authorblockA{\authorrefmark{4}
                 Dept.~of Math.,
                 Indiana University,
                 Bloomington, IN 47405, USA,
                 \texttt{jteutsch@indiana.edu}}
\authorblockA{\authorrefmark{5}
                 Dept.~of EE,
                 University of Novi Sad,
                 21000 Novi Sad, Serbia and Montenegro,
                 \texttt{dejanv@uns.ns.ac.yu}}
}

\maketitle

\begin{abstract}
  In order to understand the performance of a code under maximum-likelihood
  (ML) decoding, it is crucial to know the minimal codewords. In the context
  of linear programming (LP) decoding, it turns out to be necessary to know
  the minimal pseudo-codewords. This paper studies the minimal codewords and
  minimal pseudo-codewords of some families of codes derived from projective
  and Euclidean planes. Although our numerical results are only for codes of
  very modest length, they suggest that these code families exhibit an
  interesting property. Namely, all minimal pseudo-codewords that are not
  multiples of a minimal codeword have an AWGNC pseudo-weight that is strictly
  larger than the minimum Hamming weight of the code. This observation has
  positive consequences not only for LP decoding but also for iterative
  decoding.
\end{abstract}

\section{Introduction}
\label{sec:introduction:1}

Our motivation for looking at minimal codewords and minimal pseudo-codewords
(PCWs) is twofold. On the one hand we would like to be able to give
performance guarantees of the LP decoder, on the other hand, the connection
made by Koetter and Vontobel~\cite{Koetter:Vontobel:03:1,
Vontobel:Koetter:04:2} between iterative decoding and LP decoding suggests
that results for LP decoding have immediate implications for iterative
decoding. In this paper we focus solely on certain families of codes based on
projective and Euclidean planes. One of the reasons why these families are
worthwhile study objects is that in the past, several groups of authors have
experimentally observed that codes from these families can perform very well
under iterative decoding, see e.g.~\cite{Lucas:Fossorier:Kou:Lin:00:1,
Kou:Lin:Fossorier:01:1}.
Another reason is that these families of codes have concise descriptions and
large automorphism groups which may potentially be used to simplify their
analysis.

More precisely, the codes under investigation are the families of codes that
were called type-I PG-LDPC and type-I EG-LDPC codes
in~\cite{Kou:Lin:Fossorier:01:1}. Type-I PG-LDPC codes are defined as
follows. Let $q \defeq 2^s$ for some positive integer $s$ and consider a
(finite) projective plane $\PGq$ (see e.g.~\cite{Batten:97}) with $q^2 + q +
1$ points and $q^2 + q + 1$ lines: each point lies on $q+1$ lines and each
line contains $q+1$ points.\footnote{Note that the ``$2$'' in $\PGq$ stands
for the dimensionality of the geometry, which in the case of planes is $2$.} A
standard way of associating a parity-check matrix $\matr{H}$ of a binary
linear code to a finite geometry is to let the columns of $\matr{H}$
correspond to the set of points, to let the rows of $\matr{H}$ correspond to
the set of lines, and finally to define the entries of $\matr{H}$ according to
the incidence structure of the finite geometry. In this way, we can associate
to the projective plane $\PGq$ the code $\codePGq$ with parity-check matrix
$\matr{H} \defeq \matr{H}_{\PGq}$ whose parameters are:
\begin{center}
  \begin{tabular}{ll}
    length                                         & $n      = q^2 + q + 1$ \\
    dimension                                      & $k      = n - 3^s - 1$ \\
    minimum Hamming distance                       & $\dmins = q + 2$ \\
    uniform column weight of $\matr{H}$            & $\wcol  = q + 1$ \\
    uniform row weight of $\matr{H}$               & $\wrow  = q + 1$ \\
    size of $\matr{H}$                             & $n \times n$
  \end{tabular}
\end{center}

Type-I EG-LDPC codes are defined as follows. Let $q \defeq 2^s$ for some
positive integer $s$ and consider a (finite) Euclidean plane $\EGq$ (see
e.g.~\cite{Batten:97}) with $q^2$ points and $q^2 + q$ lines: each point lies
on $q+1$ lines and each line contains $q$ points. We essentially use the same
procedure as outlined above in order to associate a parity-check matrix to a
finite geometry. But before doing this, we modify the Euclidean plane
slightly: we select a point of $\EGq$ and remove it together with the $q+1$
lines through it. Doing so, we obtain an $\EGq$-based code $\codeEGq$ with
parity-check matrix $\matr{H} \defeq \matr{H}_{\EGq}$ whose parameters
are:
\begin{center}
  \begin{tabular}{ll}
    length                                         & $n      = q^2 - 1$ \\
    dimension                                      & $k      = n - 3^s + 1$ \\
    minimum Hamming distance                       & $\dmins = q + 1$ \\
    uniform column weight of $\matr{H}$            & $\wcol  = q$ \\
    uniform row weight of $\matr{H}$               & $\wrow  = q$ \\
    size of $\matr{H}$                             & $n \times n$
  \end{tabular}
\end{center}

Both families of codes have the nice property that with an appropriate
ordering of the columns and rows, the parity-check matrix is a circulant
matrix, meaning that $\codePGq$ and $\codeEGq$ are cyclic codes. This fact can
e.g.~be used for efficient encoding. Such symmetries can also substantially
simplify the analysis; let us point out that the automorphism groups of
$\codePGq$ and $\codeEGq$ actually contain many more automorphisms besides the
cyclic-shift automorphism implied by the cyclicity of the codes.

The structure of the rest of the paper is as
follows. Sec.~\ref{sec:ml:lp:decoding:1} discusses ML and LP decoding and
Secs.~\ref{sec:minimal:codewords:1}
and~\ref{sec:fc:minimal:pseudo:codewords:1} introduce minimal codewords and
minimal PCWs, respectively. The aim of these earlier sections is
to set the stage for Sec.~\ref{sec:numerical:results:1} and to enable the
reader to appreciate the numerical results presented therein for certain
selected codes. Finally, in Sec.~\ref{sec:conclusions:1} we state some
concluding remarks.

\section{ML and LP Decoding}
\label{sec:ml:lp:decoding:1}

In this section we briefly review ML and LP decoding. Consider a binary linear
code $\code{C}$ of length $n$ and dimension $k$ that is used for data
transmission over a memoryless binary-input channel. The codeword that is
transmitted will be called $\vx$ whereas the received vector will be called
$\vy$. Based on the received vector, we can define the log-likelihood ratios
(LLRs) to be $\lambda_i \defeq \log \big( p_{Y_i|X_i}(y_i|0) /
p_{Y_i|X_i}(y_i|1) \big)$, $i = 1, \ldots, n$. ML decoding can then be cast as
\begin{align}
  \hvx
    &\defeq 
       \arg \min_{\vx \in \code{C}}
         \sum_{i=1}^{n}
           x_i \lambda_i,
             \label{eq:ml:decoder:1}
\end{align}
Letting $\convhull(\code{C})$ be the convex hull of $\code{C}$ in $\R^n$, the
above ML decoding rule can also be formulated as
\begin{align}
  \hvx
    &\defeq
       \arg \min_{\vx \in \convhull(\code{C})}
         \sum_{i=1}^{n}
           x_i \lambda_i.
             \label{eq:ml:decoder:2}
\end{align}
Unfortunately, for most codes of interest, the description complexity of
$\convhull(\code{C})$ grows exponentially in the block length and therefore
finding the minimum in \eqref{eq:ml:decoder:2} with a linear programming
solver is highly impractical for reasonably long codes.\footnote{Exceptions to
this observation include for example the class of convolutional codes with not
too many states.}

The next step is to use a standard approach from optimization practice: we
replace the minimization over $\convhull(\code{C})$ by a minimization over
some easily describable polytope $\fp{P}$ that is a relaxation of
$\convhull(\code{C})$:
\begin{align}
  \hvx
    &\defeq \arg \min_{\vx \in \fp{P}}
       \sum_{i=1}^{n}
         x_i \lambda_i.
           \label{eq:lp:decoder:1}
\end{align}
If $\fp{P}$ is strictly larger than $\convhull(\code{C})$ then the decision
rule in~\eqref{eq:lp:decoder:1} obviously represents a sub-optimal decoder. A
relaxation that works particularly well for LDPC codes is given by the
following approach~\cite{Feldman:03:1, Feldman:Wainwright:Karger:05:1}. Let
$\code{C}$ be described by an $m \times n$ parity-check matrix $\matr{H}$ with
rows $\vect{h}_1, \vect{h}_2, \ldots, \vect{h}_m$. Then the polytope $\fp{P}
\defeq \fph{P}{H}$, in this context also called the fundamental
polytope~\cite{Koetter:Vontobel:03:1}, is defined as

\vspace{-0.3cm}
{\small
\begin{align*}
  \fp{P}
    &\defeq
       \bigcap_{i=1}^{m}
         \convhull(\code{C}_i)
  \text{ with }
  \code{C}_i
     \defeq \left\{
              \vx \in \{0, 1\}^n
              \, \left| \, 
                \vect{h}_i \vx^\tr = 0
                \, \operatorname{mod}\, 2
              \right.
            \right\}.
\end{align*}
}%
Note that $\fp{P}$ is a convex set within $[ 0, 1 ]^n$ that contains
$\convhull(\code{C})$ but whose description complexity is much smaller than
the one of $\convhull(\code{C})$. Points in the set $\fp{P}$ will be called
PCWs. Because the set $\fp{P}$ is usually strictly larger than
$\convhull(\code{C})$, it can obviously happen that the decoding rule in
\eqref{eq:lp:decoder:1} delivers a vertex of $\fp{P}$ that is not a
codeword. Such vertices are the reason for the sub-optimality of LP decoding
(cf.~\cite{Feldman:Wainwright:Karger:05:1,Koetter:Vontobel:03:1}).  Note that
$\fph{P}{H}$ is a function of the parity-check matrix $\matr{H}$ that
describes the code $\code{C}$; different parity-check matrices for the same
code might therefore lead to different fundamental polytopes.

\section{Minimal Codewords}
\label{sec:minimal:codewords:1}

Although ML decoding is often impractical, knowing bounds on the block error
rate of an ML decoder can help in assessing the performance of sub-optimal but
practical decoding algorithms. 

\begin{Definition}[cf.~e.g.~\cite{Hwang:79:1, Agrell:96:1}]
  \label{def:minimal:codewords:1}

  Let $\code{C}$ be a binary linear code of length $n$. For $\vx \in
  \code{C}$, let $\setDML{\vect{x}} \defeq \big\{ \big. \vlambda \in \R^n \
  \big| \ \vx' \cdot \vlambda^\tr \geq \vx \cdot \vlambda^\tr \text{ for all }
  \vx' \in \code{C} \setminus \{ \vx \} \big\}$ be the region in the LLR space
  where the ML decoder decides in favor of the codeword $\vx$.\footnote{Note
  that during ML decoding, ties between decoding regions can either be
  resolved in a random or in a systematic fashion.} \\ \mbox{}\edefinition
\end{Definition}

In the following, we will assume that we use a binary linear code $\code{C}$
for data transmission over a binary-input output-symmetric channel. For this
setup, when studying the ML decoder in \eqref{eq:ml:decoder:1} or
\eqref{eq:ml:decoder:2}, we can without loss of generality assume that the
zero codeword was sent, because all decision regions are congruent.

Our interest in the following definition will become apparent in
Th.~\ref{theorem:minimal:codewords:1} below.

\begin{Definition}
  \label{lemma:binary:code:minimal:codeword:properties:1:1}

  Let the \emph{support} of a vector $\vect{\vx}$ be defined as $\supp(\vx)
  \defeq \big \{ \big. i \, \big| \, x_i \neq 0 \big\}$ and let $\code{C}$ be
  a binary code. A non-zero codeword $\vx \in \code{C}$ is called
  \emph{minimal} if and only if its support does not (strictly) contain the
  support of any other non-zero codeword as a proper subset. The set of all
  minimal codewords of $\code{C}$ will be denoted by
  $\set{M}(\code{C})$. \edefinition
\end{Definition}

\begin{Theorem}[cf.~e.g.~\cite{Agrell:96:1}]
  \label{theorem:minimal:codewords:1}
  
  Let $\code{C}$ be a binary linear code of length $n$. The decision region
  $\setDML{\vect{x}}$ of a codeword $\vx \in \code{C}$ shares a facet with the
  decision region $\setDMLzero$ of the zero codeword if and only if $\vx \in
  \set{M}(\code{C})$. \etheorem
\end{Theorem}

Therefore, knowing the minimal codewords of the code $\code{C}$ is sufficient
in order to assess its ML decoding performance. (Further results about minimal
codewords can e.g.~be found in~\cite{Hwang:79:1, Agrell:96:1,
Ashikhmin:Barg:98:1, Borissov:Manev:Nikova:01:1}.)

\section{The Fundamental Cone and \\ Minimal Pseudo-Codewords}
\label{sec:fc:minimal:pseudo:codewords:1}

For LP decoding of a binary linear code that is used for data transmission
over a binary-input output-symmetric channel, it is sufficient to consider the
part of the fundamental polytope $\fp{P}$ around the vertex $\vect{0}$,
cf.~\cite{Koetter:Vontobel:03:1}. (See also ~\cite{Feldman:03:1,
Feldman:Wainwright:Karger:05:1} that discuss this so-called
``$\code{C}$-symmetry'' property.)

\begin{Definition}[\!\!\cite{Koetter:Vontobel:03:1, 
                         Feldman:Wainwright:Karger:05:1}]
  \label{def:fundamental:cone:1}

  Let $\code{C}$ be an arbitrary binary linear code and let $\matr{H}$ be its
  parity-check matrix. We let $\set{J} \defeq \set{J}(\matr{H})$ be the set of
  row indices of $\matr{H}$ and we let $\set{I} \defeq \set{I}(\matr{H})$ be
  the set of column indices of $\matr{H}$, respectively. For each $j \in
  \set{J}$, we let $\set{I}_j \defeq \set{I}_j(\matr{H}) \defeq \big\{ i \in
  \set{I} \ | \ h_{ji} = 1 \big\}$. We define the {\em fundamental cone}
  $\fch{K}{H}$ of $\matr{H}$ to be the set of vectors $\vomega \in \R^n$ that
  satisfy
  \begin{alignat*}{2}
    \forall i \in \set{I}:&&
        \quad
    \omega_i
      &\geq 0, \\
    \forall j \in \set{J}, \
      \forall i \in \set{I}_j:&&
        \quad
    \sum_{i' \in \set{I}_j \setminus \{ i \}}
      \omega_{i'}
      &\geq
        \omega_{i}.  
  \end{alignat*}\\[-0.7cm]
  \mbox{}\edefinition
\end{Definition}

The fundamental cone defined in Def.~\ref{def:fundamental:cone:1} is exactly
the part of the fundamental polytope $\fp{P}$ around the vertex $\vect{0}$ and
stretched to infinity. We note that if $\vomega \in \fch{K}{H}$, then also
$\alpha \cdot \vomega \in \fch{K}{H}$ for any $\alpha > 0$. Moreover, for any
$\vomega \in \fch{K}{H}$ there exists an $\alpha > 0$ (in fact, a whole
interval of $\alpha$'s) such that $\alpha \cdot \vomega \in \fph{P}{H}$.

  For a given binary linear code $\code{C}$ with parity-check matrix
  $\matr{H}$, the importance of the set $\fch{K}{H}$ lies in the following
  fact. Let $\setDLPzero \defeq \big\{ \big. \vlambda \in \R^n \ \big| \
  \vomega \cdot \vlambda^\tr \geq 0 \text{ for all } \vomega \in \fph{P}{H}
  \setminus \{ \vect{0} \} \big\}$ be the region where the LP decoder decides
  in favor of the codeword $\vect{0}$.\footnote{Note that during LP decoding,
  ties between decoding regions can either be resolved in a random or in a
  systematic fashion.} It can easily be seen that $\setDLPzero = \big\{
  \big. \vlambda \in \R^n \ \big| \ \vomega \cdot \vlambda^\tr \geq 0
  \text{ for all } \vomega \in \fch{K}{H} \setminus \{ \vect{0} \}
  \big\}$. Therefore, when studying LP decoding it is enough to know
  $\fch{K}{H}$; all vectors $\vomega \in \fch{K}{H}$ will henceforth be called
  PCWs.

\begin{Definition}[\!\!\cite{Koetter:Vontobel:03:1}]
  \label{def:minimal:pseudo:codewor:1}

  Let $\code{C}$ be an arbitrary binary linear code described by the
  parity-check matrix $\matr{H}$ whose fundamental cone is $\fch{K}{H}$. A
  vector $\vomega \in \fch{K}{H}$ is called a \emph{minimal PCW}
  if the set $\{ \alpha \cdot \vomega \ | \ \alpha \geq 0 \}$ is an edge of
  $\fch{K}{H}$. Moreover, the set of all minimal PCWs will be
  called $\Mps(\fch{K}{H})$.\footnote{Note that this definition implies that
  $\vect{0} \notin \Mps(\fch{K}{H})$.} \edefinition
\end{Definition}

  For a given binary linear code $\code{C}$ with parity-check matrix
  $\matr{H}$, the importance of the set $\Mps(\fch{K}{H})$ lies in the
  following fact. From basic cone properties
  (cf.~e.g.~\cite{Boyd:Vandenberghe:04:1}), it can easily be seen that
  $\setDLPzero = \big\{ \big. \vlambda \in \R^n \ \big| \ \vomega \cdot
  \vlambda^\tr \geq 0 \text{ for all } \vomega \in \Mps(\fch{K}{H})
  \big\}$. Therefore, the set $\Mps(\fch{K}{H})$ completely characterizes the
  behavior of the LP decoder.

\begin{Definition}
  Let $\code{C}$ be an arbitrary binary linear code described by the
  parity-check matrix $\matr{H}$. The additive white Gaussian noise channel
  (AWGNC) pseudo-weight~\cite{Forney:Koetter:Kschischang:Reznik:01:1} of a PCW
  $\vomega \in \fch{K}{H}$ is defined to be $\wpsAWGNC(\vomega) =
  ||\vomega||_1^2 / ||\vomega||_2^2$, where $||\vomega||_1$ and
  $||\vomega||_2$ are the $L_1$- and $L_2$- norm of $\vomega$,
  respectively.\footnote{We set $\wpsAWGNC(\vx) \defeq 0$ for $\vx =
  \vect{0}$. Note that for $\vx \in \{ 0, 1 \}^n$ we have $\wpsAWGNC(\vx) =
  \wH(\vx)$, where $\wH(\vx)$ is the Hamming weight of $\vx$.} \edefinition
\end{Definition}

The significance of $\wpsAWGNC(\vomega)$ is the following: it can be shown
that the squared Euclidean distance from the point $+\vect{1}$ in signal space
(which corresponds to the codeword $\vect{0}$) to the plane $\big\{ \vlambda
\in \R^n \ | \ \vomega \cdot \vlambda^\tr = 0 \big\}$ is $\wpsAWGNC(\vomega)$.

\begin{Definition}
  Let $\code{C}$ be an arbitrary binary linear code. We define the
  \emph{codeword weight enumerator} and the \emph{minimal codeword weight
  enumerator} to be the polynomials

  \vspace{-0.3cm}
  {\small
  \begin{align*}
    \chicw{\code{C}}(X)
      &\defeq
         \sum_{\vx \in \code{C}}
           X^{\wH(\vx)}
    \ \ \text{ and }\ \ 
    \chimcw{\code{C}}(X)
       \defeq
         \sum_{\vx \in \set{M}(\code{C})}
           X^{\wH(\vx)},
  \end{align*}
  }%
  respectively. \edefinition
\end{Definition}

\begin{Definition}
  Let $\code{C}$ be an arbitrary binary linear code described by the
  parity-check matrix $\matr{H}$. We define the \emph{minimal PCW
  AWGNC pseudo-weight enumerator} to be the polynomial (with potentially
  non-integer exponents)

  \vspace{-0.3cm}
  {\small
  \begin{align*}
    \chiAWGNCmpcw{\matr{H}}(X)
      &= \sum_{[\vomega] \in \Mps(\fch{K}{H})}
           X^{\wpsAWGNC(\vomega)},
  \end{align*}
  }%
  where the summation is over all equivalence classes of minimal
  PCWs.\footnote{Two PCWs $\vomega, \vomega' \in
  \fch{K}{H}$ are in the same equivalence class if there exists an $\alpha >
  0$ such that $\vomega = \alpha \cdot \vomega'$.} \edefinition
\end{Definition}

\begin{Definition}
  Let $\code{C}$ be an arbitrary binary linear code described by the
  parity-check matrix $\matr{H}$ and let $\Mps'(\fch{K}{H})$ be the set of all
  minimal PCWs that are \emph{not} multiples of minimal codewords.
  We call the real-valued quantity

  \vspace{-0.3cm}
  {\small
  \begin{align*}
    g(\matr{H})
      &\defeq \min_{\vomega \in \Mps'(\fch{K}{H})} \wpsAWGNC(\vomega)
       -
       \wHmin(\code{C}(\matr{H}))
  \end{align*}
  }
  the pseudo-weight spectrum gap of $\matr{H}$.\edefinition
\end{Definition}

  Using Cor.~8 in~\cite{Koetter:Vontobel:03:1} one can show that for a
  randomly constructed $(\wcol,\wrow)$-regular code with $3 \leq \wcol <
  \wrow$ the pseudo-weight spectrum gap becomes \emph{strictly negative} with
  probability one as the block length goes to infinity. However, using Th.~1
  in~\cite{Vontobel:Koetter:04:1} one can show that for the $\PGq$- and
  $\EGq$-based codes (with square parity-check matrix as discussed in
  Sec.~\ref{sec:introduction:1}) the pseudo-weight spectrum gap is
  \emph{non-negative}. In fact, we will see that for the codes investigated in
  Sec.~\ref{sec:numerical:results:1} the pseudo-weight spectrum gap is
  significantly positive. We note that by applying simple performance bounding
  techniques it can be shown that the larger the gap is, the closer is the LP
  decoding performance (and potentially also the iterative decoding
  performance~\cite{Koetter:Vontobel:03:1}) to the ML decoding performance as
  the SNR goes to infinity.

Before we turn to some numerical results about minimal codewords and minimal
PCWs, let us mention some related work by Kashyap and
Vardy~\cite{Kashyap:Vardy:03:1} which discusses results that characterize
(minimal) stopping sets for finite-geometry-based codes. This is of some
relevance to this paper because it is well-known that the support set of any
PCW is a stopping set and that for any stopping set there exists a PCW whose
support set equals that stopping set \cite{Koetter:Vontobel:03:1,
Feldman:Wainwright:Karger:05:1}.

\section{Numerical Results}
\label{sec:numerical:results:1}

In this section we present minimal PCWs, weight enumerators, and
the pseudo-weight spectrum gap for some short $\PGq$- and $\EGq$-based codes.

\subsection{Type-I PG-LDPC code for $q = 2$}

The $\PG{2}{2}$-based code $\codePG{2}{2}$ of type I has parameters $[n {=} 7,
k {=} 3, \dmins {=} 4]$ and can be represented by the following circulant
parity-check matrix of size $7 \times 7$:

\vspace{-0.2cm}
{\scriptsize
\begin{align*}
  \matr{H}_{\PG{2}{2}}
    &= \begin{pmatrix}
         1 & 1 & 0 & 1 & 0 & 0 & 0 \\
         0 & 1 & 1 & 0 & 1 & 0 & 0 \\
         0 & 0 & 1 & 1 & 0 & 1 & 0 \\
         0 & 0 & 0 & 1 & 1 & 0 & 1 \\
         1 & 0 & 0 & 0 & 1 & 1 & 0 \\
         0 & 1 & 0 & 0 & 0 & 1 & 1 \\
         1 & 0 & 1 & 0 & 0 & 0 & 1
       \end{pmatrix}.
\end{align*}
}%
The set $\set{M}(\code{C})$ of minimal codewords consists of the following
codewords:

\vspace{-0.3cm}
{\scriptsize
\begin{align*}
  &
   (1, 0, 0, 1, 0, 1, 1),\ 
   (1, 1, 0, 0, 1, 0, 1),\ 
   (1, 1, 1, 0, 0, 1, 0),\
   (0, 1, 1, 1, 0, 0, 1), \\
 &
   (1, 0, 1, 1, 1, 0, 0),\
   (0, 1, 0, 1, 1, 1, 0),\
   (0, 0, 1, 0, 1, 1, 1).
\end{align*}
}%
Obviously, all of them have Hamming weight $4$ and they are all cyclic shifts
of each other. Because the code has $2^3 = 8$ codewords in total, it turns
that this code is special in the following sense: there are no non-zero
codewords that are not minimal codewords.

The set $\Mps(\fch{K}{H})$ of minimal codewords contains all the elements from
$\set{M}(\code{C})$ plus the following PCWs that are not codewords
(we show one representative per equivalence class):

\vspace{-0.3cm}
{\scriptsize
\begin{align*}
  &
    (1, 2, 2, 1, 2, 1, 1),\
    (1, 1, 2, 2, 1, 2, 1),\
    (1, 1, 1, 2, 2, 1, 2),\
    (2, 1, 1, 1, 2, 2, 1), \\
  &
    (1, 2, 1, 1, 1, 2, 2),\
    (2, 1, 2, 1, 1, 1, 2),\
    (2, 2, 1, 2, 1, 1, 1).
\end{align*}
}%
All these minimal PCWs that are not codewords turn out to be
cyclic shifts of each other and to have AWGNC pseudo-weight $\frac{100}{16} =
6.25$. The weight enumerators for this code are therefore:

\vspace{-0.3cm}
{\small
\begin{align*}
  \chicw{\codePG{2}{2}}(X)
    &= X^0 + 7 X^4, \\
  \chimcw{\codePG{2}{2}}(X)
    &= 7 X^4, \\
  \chiAWGNCmpcw{\matr{H}_{\PG{2}{2}}}(X)
    &= 7 X^4 + 7 X^{6.25}.
\end{align*}
}%
Hence, the pseudo-weight spectrum gap is $g(\matr{H}_{\PG{2}{2}}) = 6.25 - 4 =
2.25$.

The codes introduced in Sec.~\ref{sec:introduction:1} were based on square
parity-check matrices. However, the code $\PG{2}{2}$ can also be described by
a parity-check matrix $\matr{H}'_{\PG{2}{2}}$ of size $4 \times 7$ which is
equivalent to the first four lines of the $\matr{H}_{\PG{2}{2}}$. The minimal
PCWs that are not codewords turn out to be (we show one
representative per equivalence class):

\vspace{-0.3cm}
{\scriptsize
\begin{align*}
  &
    (3, 2, 1, 1, 1, 0, 0),\ 
    (0, 1, 2, 1, 1, 3, 0),\ 
    (0, 1, 1, 1, 2, 0, 3),\
    (0, 1, 1, 1, 1, 0, 0), \\
  &
    (2, 1, 1, 1, 0, 0, 1),\
    (2, 1, 0, 1, 1, 1, 0),\
    (1, 2, 1, 1, 1, 0, 0),\
    (0, 1, 2, 1, 1, 1, 0), \\
  &
    (0, 1, 1, 1, 2, 0, 1),\
    (0, 1, 1, 1, 0, 2, 1),\
    (1, 0, 1, 1, 1, 2, 0),\
    (1, 0, 1, 1, 1, 0, 2), \\
  &
    (0, 1, 0, 1, 1, 1, 2),  
\end{align*}
}%
from which follows that

\vspace{-0.3cm}
{\small
\begin{align*}
  \chiAWGNCmpcw{\matr{H}'_{\PG{2}{2}}}(X)
    &= 11 X^4 + 9 X^{4.5}.
\end{align*}
}%
Note that the pseudo-weight spectrum gap is $g(\matr{H}'_{\PG{2}{2}}) = 4 - 4
= 0$. Comparing the enumerator $\chiAWGNCmpcw{\matr{H}_{\PG{2}{2}}}(X)$ with
the enumerator $\chiAWGNCmpcw{\matr{H}'_{\PG{2}{2}}}(X)$ it is apparent that
the performance of LP decoding using the second representation will be worse
than the performance of LP decoding using the first representation. Based on
iterative decoder simulations, MacKay and
Davey~\cite[Sec.~4]{MacKay:Davey:01:1} observed a similar performance
hierarchy between different representations of the same code. (Note that the
code under investigation in~\cite{MacKay:Davey:01:1} was the
$\PG{2}{16}$-based code.)

\subsection{Type-I PG-LDPC code for $q = 4$}

\begin{figure}
  \begin{center}
    \epsfig{file=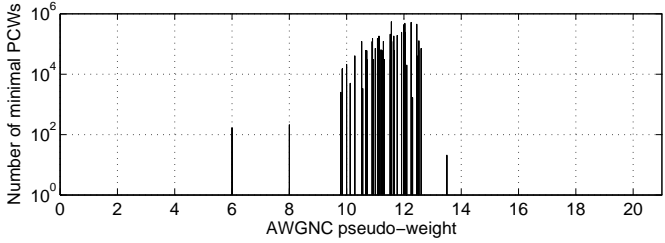,
            width=\linewidth}
  \end{center}
  \caption{Histogram of the AWGNC pseudo-weight of minimal
  PCWs of the $\PG{2}{4}$-based code. (Note that the y-axis is
  logarithmic.)}
  \label{fig:pg:code:q:4:min:pseudo:codewords:histogram:1}
\end{figure}

The parity-check matrix $\matr{H}_{\PG{2}{4}}$ of the $\PG{2}{4}$-based code
$\codePG{2}{4}$ has size $21 \times 21$, uniform column and row weight $5$,
and yields a code with parameters $[n {=} 21, k {=} 11, \dmins {=} 6]$. The
codeword weight enumerator and the minimal codeword weight enumerator are

\vspace{-0.3cm}
{\small
\begin{align*}
  \chicw{\codePG{2}{4}}(X)
    &= X^0 + 168 X^6 + 210 X^8 + 1008 X^{10} \\
    &\quad\
       + 280 X^{12} + 360 X^{14} + 21 X^{16}, \\
  \chimcw{\codePG{2}{4}}(X)
    &= 168 X^6 + 210 X^8 + 1008 X^{10},
\end{align*}
}%
respectively. Looking at these enumerators we see that all codewords with
Hamming weight $6$, $8$, and $10$ are also minimal codewords. Analyzing the
set of all weight-$6$ codewords one sees that they all have the same pattern,
i.e.~they can all be obtained from a single weight-$6$ codeword by applying a
suitable $\PG{2}{4}$-automorphism. The same is true for all other sets of
codewords with the same weight. This makes the classification of all the
codewords of $\codePG{2}{4}$, and in particular of the minimal codewords of
$\codePG{2}{4}$, relatively easy.

Instead of giving the formula for $\chiAWGNCmpcw{\matr{H}_{\PG{2}{4}}}(X)$, we
simply give its histogram (spectrum),
cf.~Fig.~\ref{fig:pg:code:q:4:min:pseudo:codewords:histogram:1}. The gap turns
out to be $g(\matr{H}_{\PG{2}{4}}) = 9.8 - 6 = 3.8$.

\subsection{Type-I PG-LDPC code for $q = 8$}

Judging from some very preliminary results (based on random search
experiments) the pseudo-weight spectrum gap $g(\matr{H}_{\PG{2}{8}})$ for the
$\PG{2}{8}$-based code seems to be at least~$6.0$.

\subsection{Type-I EG-LDPC code for $q = 4$}

The parity-check matrix $\matr{H}_{\EG{2}{4}}$ of the $\EG{2}{4}$-based code
$\codeEG{2}{4}$ has size $15 \times 15$, uniform column and row weight $4$,
and yields a code with parameters $[n {=} 15, k {=} 7, \dmins {=} 5]$. The
codeword weight enumerator and the minimal codeword weight enumerator are

\vspace{-0.3cm}
{\small
\begin{align*}
  \chicw{\codeEG{2}{4}}(X)
    &=  X^0 + 18 X^5 + 30 X^6 + 15 X^7 \\
    &\quad\
        + 15 X^8 + 30 X^9 + 18 X^{10} + X^{15}\\
  \chimcw{\codeEG{2}{4}}(X)
    &=  18 X^5 + 30 X^6 + 15 X^7 + 15 X^8 + 30 X^9,
\end{align*}
}%
respectively. Looking at these enumerators we see that all codewords with
Hamming weight $5$, $6$, $7$, $8$, and $9$ are also minimal
codewords. Analyzing the set of all weight-$5$ codewords one sees that they
all have the same pattern, i.e.~they can all be obtained from a single
weight-$5$ codeword by applying a suitable $\EG{2}{4}$-automorphism. The same
is true for all other sets of codewords with the same weight.

\begin{figure}
  \begin{center}
    \epsfig{file=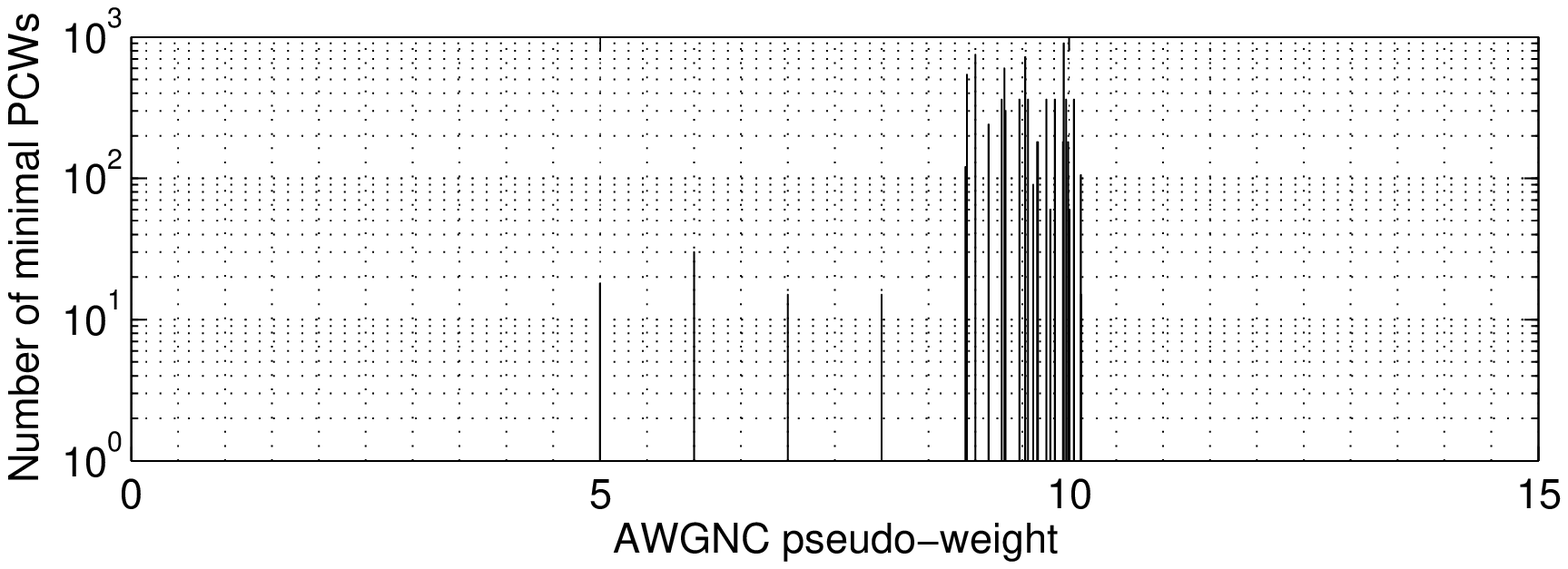,
            width=\linewidth}
    \epsfig{file=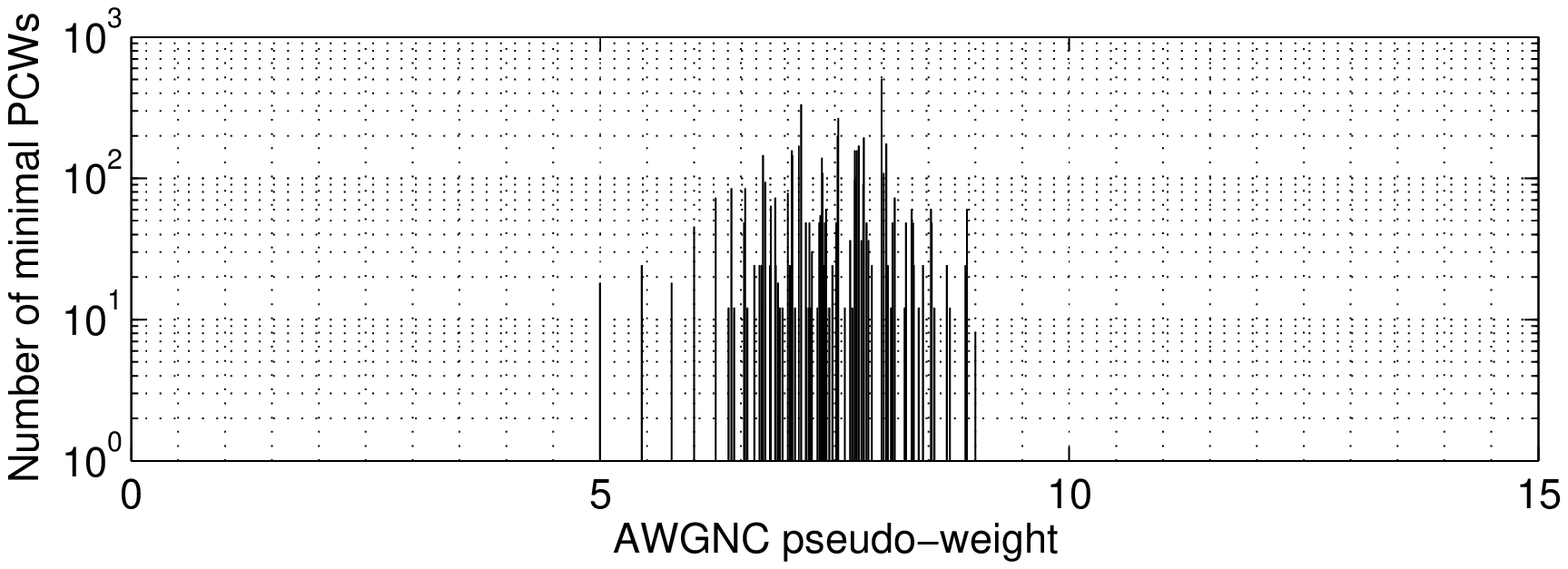,
            width=\linewidth}
    \epsfig{file=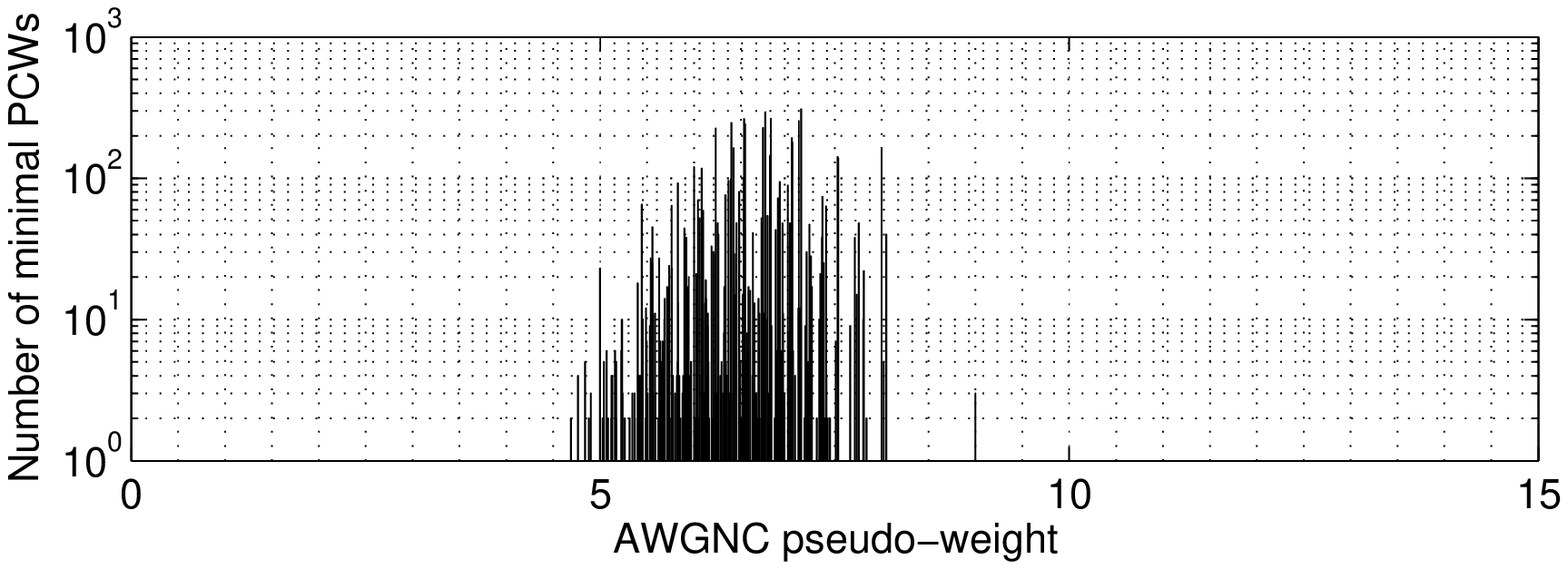,
            width=\linewidth}
  \end{center}
  \caption{Histogram of the AWGNC pseudo-weight of minimal PCWs of
  the $\EG{2}{4}$-based code. (Note that the y-axis is logarithmic.) Top: For
  $15 \times 15$ parity-check matrix $\matr{H}_{\EG{2}{4}}$,
  $g(\matr{H}_{\EG{2}{4}}) = \frac{169}{19} - 5 \approx 8.89 - 5 =
  3.89$. Middle: For $9 \times 15$ parity-check matrix
  $\matr{H}'_{\EG{2}{4}}$, $g(\matr{H}'_{\EG{2}{4}}) = \frac{49}{9} - 5
  \approx 5.44 - 5 = 0.44$. Bottom: For $8 \times 15$ parity-check matrix
  $\matr{H}''_{\EG{2}{4}}$, $g(\matr{H}''_{\EG{2}{4}}) = \frac{361}{77} - 5
  \approx 4.69 - 5 = -0.31$.}
  \label{fig:eg:code:q:4:min:pseudo:codewords:histogram:1}
\end{figure}

\begin{figure}
  \begin{center}
    \epsfig{file=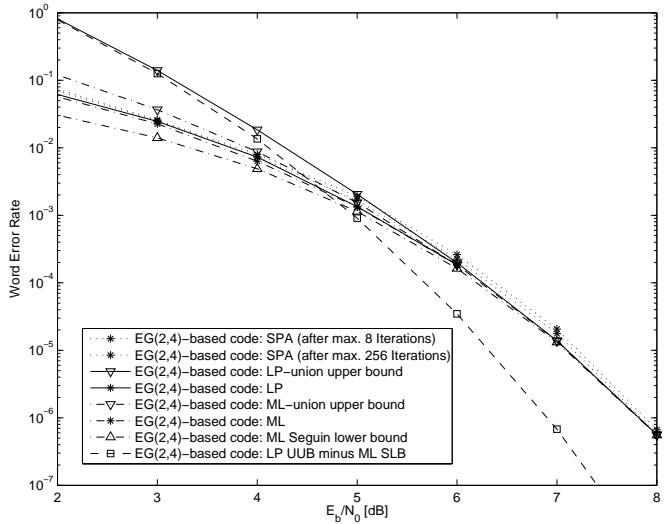,
            width=\linewidth}
  \end{center}
  \caption{Word error rate for various decoding algorithms together with some
    upper and lower bounds. (See main text for explanations.)}
  \label{fig:eg:code:q:4:decoding:and:bounds:3}
\end{figure}

The histograms (spectra) in
Fig.~\ref{fig:eg:code:q:4:min:pseudo:codewords:histogram:1} correspond to
various parity-check matrices that describe
$\codeEG{2}{4}$. Fig.~\ref{fig:eg:code:q:4:min:pseudo:codewords:histogram:1}
(top) shows the histogram for $\chiAWGNCmpcw{\matr{H}_{\PG{2}{4}}}(X)$;
Fig.~\ref{fig:eg:code:q:4:min:pseudo:codewords:histogram:1} (middle) shows the
histogram for $\chiAWGNCmpcw{\matr{H}'_{\PG{2}{4}}}(X)$ where
$\matr{H}'_{\PG{2}{4}}$ is a randomly selected $9 \times 15$ submatrix (with
column weights at least $2$) of $\matr{H}_{\PG{2}{4}}$; and finally
Fig.~\ref{fig:eg:code:q:4:min:pseudo:codewords:histogram:1} (bottom) shows the
histogram for $\chiAWGNCmpcw{\matr{H}''_{\PG{2}{4}}}(X)$ where
$\matr{H}''_{\PG{2}{4}}$ is an $8 \times 15$ submatrix (with five columns
having weight only one) of consecutive rows of the (circulant) matrix
$\matr{H}_{\PG{2}{4}}$. It can easily be seen that for the parity-check
matrices under investigation those with more lines lead to more favorable
histograms.

In Fig.~\ref{fig:eg:code:q:4:decoding:and:bounds:3} we show various decoding
simulation results for data transmission over a binary-input AWGNC and lower
and upper bounds: $\matr{H}_{\EG{2}{4}}$-based sum-product algorithm decoding,
$\matr{H}_{\EG{2}{4}}$-based LP decoding, $\codeEG{2}{4}$-based ML decoding,
an upper bound on LP decoding based on a union of events upper bound, an upper
bound on ML decoding based on a union of events upper bound, and a lower bound
on ML decoding based on an inequality by de Caen as presented by
S\'eguin~\cite{Seguin:98:1}. It can be seen that thanks to the knowledge of
minimal codewords and minimal PCWs we are able to obtain bounds
that are very tight from a certain SNR value on.

\subsection{How the results were obtained}

Let us briefly mention how the results for the minimal PCWs were
obtained. We used the program ``lrs''~\cite{Avis:00:1} to search edges in
cones. For the code $\codePG{2}{4}$ we additionally used the two-transitivity
of the points of a projective plane in order to formulate a simpler
edge-enumeration subproblem which can be solved efficiently and from which all
the minimal PCWs can be derived. There are various other ways to
use the large automorphism groups of these codes that help in simplifying the
edge-enumerating problem. Properties of minimal codewords might also be used
towards that goal.

\section{Concluding Remarks}
\label{sec:conclusions:1}

We have investigated the minimal PCWs of some simple $\PGq$- and
$\EGq$-based binary linear codes and we have introduced the notion of a
pseudo-weight spectrum gap for a parity-check matrix, a concept which is
certainly worthwhile to be further explored. Although our numerical results
are for codes of very modest size, to the best of our knowledge this is the
first study that tries to \emph{analytically} quantify the behavior of $\PGq$-
and $\EGq$-based binary linear codes under LP and iterative
decoding. Extending these results to somewhat longer codes has the potential
to explain many experimental observations made in the past.

\section*{Acknowledgment}

This paper is based on work started at the IMA Participating Institutions
Summer Program for graduate students on ``Coding and Cryptography'' at Notre
Dame University, Notre Dame, IN, USA, June 7-25, 2004. We would like to take
this opportunity to thank the organizers. P.O.V.'s research was supported by
NSF Grants CCR 99-84515, CCR 01-05719, and ATM-0296033 and by DOE SciDAC and
ONR Grant N00014-00-1-0966.


\begin{thebibliography}{10}

{\scriptsize

\bibitem{Koetter:Vontobel:03:1}
R.~Koetter and P.~O. Vontobel, ``Graph covers and iterative decoding of
  finite-length codes,'' in {\em Proc.\ 3rd Intern.~Conf.~on Turbo Codes and
  Related Topics}, (Brest, France), pp.~75--82, Sept.~1--5 2003.

\bibitem{Vontobel:Koetter:04:2}
P.~O. Vontobel and R.~Koetter, ``On the relationship between linear programming
  decoding and min-sum algorithm decoding,'' in {\em Proc.\ Intern.\ Symp.\ on
  Inform.\ Theory and its Applications (ISITA)}, (Parma, Italy), pp.~991--996,
  Oct.~10--13 2004.

\bibitem{Lucas:Fossorier:Kou:Lin:00:1}
R.~Lucas, M.~Fossorier, Y.~Kou, and S.~Lin, ``Iterative decoding of one-step
  majority logic decodable codes based on belief propagation,'' {\em IEEE
  Trans.\ on Comm.}, vol.~COMM-48, pp.~931--937, June 2000.

\bibitem{Kou:Lin:Fossorier:01:1}
Y.~Kou, S.~Lin, and M.~P.~C. Fossorier, ``Low-density parity-check codes based
  on finite geometries: a rediscovery and new results,'' {\em IEEE Trans.\ on
  Inform.\ Theory}, vol.~IT--47, pp.~2711--2736, Nov. 2001.

\bibitem{Batten:97}
L.~M. Batten, {\em Combinatorics of Finite Geometries}.
\newblock Cambridge: Cambridge University Press, second~ed., 1997.

\bibitem{Feldman:03:1}
J.~Feldman, {\em Decoding Error-Correcting Codes via Linear Programming}.
\newblock PhD thesis, Massachusetts Institute of Technology, Cambridge, MA,
  2003.
\newblock Available online under
  \verb+http://www.columbia.edu/~jf2189/pubs.html+.

\bibitem{Feldman:Wainwright:Karger:05:1}
J.~Feldman, M.~J. Wainwright, and D.~R. Karger, ``Using linear programming to
  decode binary linear codes,'' {\em IEEE Trans.\ on Inform.\ Theory},
  vol.~IT--51, no.~3, pp.~954--972, 2005.

\bibitem{Hwang:79:1}
T.~Y. Hwang, ``Decoding linear block codes for minimizing word error rate,''
  {\em IEEE Trans.\ on Inform.\ Theory}, vol.~25, no.~6, pp.~733--737, 1979.

\bibitem{Agrell:96:1}
E.~Agrell, ``Vorono\u\i\ regions for binary linear block codes,'' {\em IEEE
  Trans.\ on Inform.\ Theory}, vol.~42, no.~1, pp.~310--316, 1996.

\bibitem{Ashikhmin:Barg:98:1}
A.~Ashikhmin and A.~Barg, ``Minimal vectors in linear codes,'' {\em IEEE
  Trans.\ on Inform.\ Theory}, vol.~44, no.~5, pp.~2010--2017, 1998.

\bibitem{Borissov:Manev:Nikova:01:1}
Y.~Borissov, N.~Manev, and S.~Nikova, ``On the non-minimal codewords in the
  binary {R}eed-{M}uller code,'' in {\em Proc.\ IEEE Intern.\ Symp.\ on
  Inform.\ Theory}, (Washington, D.C., USA), p.~39, June 24-29 2001.

\bibitem{Boyd:Vandenberghe:04:1}
S.~Boyd and L.~Vandenberghe, {\em Convex Optimization}.
\newblock Cambridge, UK: Cambridge University Press, 2004.

\bibitem{Forney:Koetter:Kschischang:Reznik:01:1}
G.~D. {Forney, Jr.}, R.~Koetter, F.~R. Kschischang, and A.~Reznik, ``On the
  effective weights of pseudocodewords for codes defined on graphs with
  cycles,'' in {\em Codes, Systems, and Graphical Models (Minneapolis, MN,
  1999)} (B.~Marcus and J.~Rosenthal, eds.), vol.~123 of {\em IMA Vol. Math.
  Appl.}, pp.~101--112, Springer Verlag, New York, Inc., 2001.

\bibitem{Vontobel:Koetter:04:1}
P.~O. Vontobel and R.~Koetter, ``Lower bounds on the minimum pseudo-weight of
  linear codes,'' in {\em Proc.\ IEEE Intern.\ Symp.\ on Inform.\ Theory},
  (Chicago, IL, USA), p.~70, June 27--July 2 2004.

\bibitem{Kashyap:Vardy:03:1}
N.~Kashyap and A.~Vardy, ``Stopping sets in codes from designs,'' in {\em
  Proc.\ IEEE Intern.\ Symp.\ on Inform.\ Theory}, (Pacifico Yokohama, Japan),
  p.~122, June 29 -- July 4 2003.

\bibitem{MacKay:Davey:01:1}
D.~J.~C. MacKay and M.~C. Davey, ``Evaluation of {G}allager codes for short
  block length and high rate applications,'' in {\em Codes, Systems, and
  Graphical Models (Minneapolis, MN, 1999)} (B.~Marcus and J.~Rosenthal, eds.),
  pp.~113--130, Springer Verlag, New York, Inc., 2001.

\bibitem{Seguin:98:1}
G.~E. S{\'e}guin, ``A lower bound on the error probability for signals in white
  {G}aussian noise,'' {\em IEEE Trans.\ on Inform.\ Theory}, vol.~44, no.~7,
  pp.~3168--3175, 1998.

\bibitem{Avis:00:1}
D.~Avis, ``lrs: A revised implementation of the reverse search vertex
  enumeration algorithm,'' in {\em Polytopes -- Combinatorics and Computation}
  (G.~Kalai and G.~M. Ziegler, eds.), pp.~177--198, Birkh{\"a}user-Verlag,
  2000.
\newblock Programs are available online under
  \verb+http://cgm.cs.mcgill.ca/~avis/C/lrs.html+.

}

\end{thebibliography}
\end{document}